\documentclass[twocolumn,showpacs,preprintnumbers,amsmath,amssymb]{revtex4}
\usepackage{CJK}
\usepackage{graphicx}% Include figure files
\usepackage{dcolumn}% Align table columns on decimal point
\usepackage{bm}% bold math

\def\btbl{\begin{tabular}} \def\etbl{\end{tabular}}
\def\bcc{\begin{center}} \def\ecc{\end{center}}
\def\beq{\begin{equation}} \def\eeq{\end{equation}}
\def\btbl{\begin{tabular}} \def\etbl{\end{tabular}}

\def\E941{{\footnotesize E941}} \def\E864{{\footnotesize E864}}
\def\NA49{{\footnotesize NA49}} \def\NA35{{\footnotesize NA35}}

\begin{document}
\title{Gluon saturation and baryon stopping in the SPS, RHIC, and LHC energy
regions }

\author{Shuang Li$^1$}
\author{Sheng-Qin ~Feng$^{1,2,3}$}

\affiliation{$^1$College of Science, China Three Gorges University,
Yichang 443002, China} \affiliation{$^2$Key Laboratory of
Quark and Lepton Physics (Huazhong Normal Univer.), Ministry of
Education£¬Wuhan 430079£¬China} \affiliation{$^3$School of Physics
and Technology, Wuhan University, Wuhan 430072, China}

\begin{abstract}
A new geometrical scaling method with gluon saturation rapidity
limit is proposed to study the gluon saturation feature of central
rapidity region of relativistic nuclear collisions. The net-baryon
number is essentially transported by valence quarks that probe the
saturation regime in the target by multiple scattering.  We take
advantage of the gluon saturation model with geometric scaling of
rapidity limit to investigate the net baryon distributions, nuclear
stopping power and gluon saturation feature in the SPS and RHIC
energy regions. Predications for the net-baryon rapidity
distributions, the mean rapidity loss and gluon saturation feature
in central Pb + Pb collisions at LHC are made in this paper.\\

\vskip0.2cm \noindent Keywowds: gluon saturation, geometrical scaling, net-baryon distributions
\end{abstract}

 \maketitle

\section{Introduction}
During the relativistic heavy-ion collisions, the fast valence
quarks in one nucleus scatter in the other nucleus by exchanging
soft gluons, leading to their redistribution in rapidity space. The
net-baryon number is essentially transported by valence quarks that
probe the saturation regime in the target by multiple scattering.

Experimental heavy-ion investigations at the Large Hadron
Collider(LHC) pay much more attention to the mid-rapidity region
since ALICE covers rapidity up to $\left| \emph{y}\right|=2$. It
provides measurements of lower $x$ than before at energy scale that
is high enough to provide crucial tests of gluon saturation.
Therefore LHC will provide more opportunities to study the gluon
saturation feature at small $x$.

At very high energies or small values of Bjorken variable $x$, the
density of partons per unit transverse area, in nucleon or nucleus
becomes so large that it would lead to a gluon saturation of
partonic distribution. The existence of this phenomenon was
confirmed in the experiments at HERA~\cite{Breitweg,Iancu}. The
typical results from the experiments contain two parts: the small
$x$ problem and the geometric scaling. It was predicted by an
effective theory, the CGC (Color Glass Condensate), which describes
the behavior of the small $x$ components of the hadronic wave
function in QCD. In this paper, we use this CGC theory to discuss
the questions of nuclear stopping and gluon saturation in
relativistic heavy-ion collisions at SPS, RHIC and LHC.

The kinetic energy of the relativistic heavy-ion collisions that is
removed from the beam and which is available for the production of a
state such as the QGP (quark gluon plasma) depends on the amount of
stopping between the colliding ions. The stopping can be estimated
from the rapidity loss experienced by the baryons in the colliding
nuclei~\cite{Arsene,Bearden1,Klay,Bearden2,Barrette,Ahle,Appelshauser}.
If the incoming beam baryons have rapidity, $y_{b}$ relative to the
CM(which has $y=0$) and average rapidity

\vspace{-0.3cm}

\begin{equation}  %%% Eq.1
      <y>=\int_0^{y_{b}}ydydN/dy/\int_0^{y_{b}}dydN/dy
\label{eq:eq1} %Eq.1
\end{equation}

\noindent the average rapidity loss is

\vspace{-0.3cm}

\begin{equation}  %%% Eq.2
      <\delta y>=y_{b}-<y>
\label{eq:eq2} %Eq.2
\end{equation}

Here \emph{dN/dy} denotes the number of net-baryons (number of
baryons minus the number of anti-baryons) per unit of rapidity. The
studies of net baryon distributions and nuclear stopping have been
discussed by non-uniform collective flow
model~\cite{Feng1,Feng2,Feng3,Feng4}. We will use gluon saturation
to study net-baryon in this paper.

Here we should mention a novel gluon saturation model proposed by
Yacine Mehtar-Tani and Georg
Wolschin~\cite{Mehtar-Tani1,Mehtar-Tani2}. An analytical scaling law
is derived within the color glass condensate framework based on
small-coupling QCD in this model.  Inspired by this
model~\cite{Mehtar-Tani1,Mehtar-Tani2}, we study the net baryon
distributions of central collisions in the SPS and RHIC energy
regions by introducing the effective quark mass and rapidity limit
of gluon saturation region. Here we use a new geometric scaling
method to define the rapidity limit of gluon saturation region. The
important difference is the definition of geometric scaling between
our model and Ref.[14,15]. It
realized~\cite{Mehtar-Tani1,Mehtar-Tani2}that the geometric scaling
is mainly about the peak positions of net-baryon rapidity
distribution. But our model realizes the geometric scaling of
net-baryon rapidity distribution is mainly about the gluon
saturation rapidity limit. The net-baryon rapidity distributions and
the mean rapidity loss in central Pb + Pb collisions at LHC are
predicted in this paper.

\section{Gluon saturation model with geometric scaling}
The ideas for the color glass condensate are motivated by HERA data
on the gluon distribution function~\cite{Breitweg}. The gluon density,
$xG(x,Q^{2})$, rises rapidly as a function of decreasing fractional
momentum, $x$, or increasing resolution $Q$. This rise in the gluon
density ultimately owes its origin to the non-Abelian nature of QCD
and that the gluons carry color charge. At higher and higher
energies, smaller $x$ and larger $Q$ become kinematics accessible.
The rapid rise with $ln(1/x)$ was expected in a variety of
theoretical
works~\cite{Mehtar-Tani1,Mehtar-Tani2,Gribov,Mueller,McLerran,Kharzeev1,Kharzeev2,Kharzeev3,Baier,Triantafyllopoulos}.
Due to the intrinsic non-linearity of QCD, gluon showers generate
more gluon showers producing an exponential avalanche toward small
$x$. The physical consequence of this exponential growth is that the
density of gluons per unit area per unit rapidity of any hadrons
including nuclei must increase rapidly as $x$
decreases~\cite{Mehtar-Tani1,Mehtar-Tani2}.

The net-baryon number of relativistic heavy ion collisions is
essentially transported by valence quarks that probe the saturation
regime in the target by multiple scatterings. During the
relativistic heavy ion collisions, the fast valence quarks in one
nucleus scatter in the other nucleus by exchanging soft gluons,
leading to their redistribution in rapidity space. The valence quark
parton distribution at large $x$ is well known, which corresponds to
the forward and backward rapidity region, to access the gluon
distribution at small $x$ in the target nucleus. Therefore, this
picture provides a clean probe of the gluon distribution, $\varphi
(x,p_{T})$ , at small $x$ in the saturation regime.

For symmetric heavy ion collisions, the contribution of the
fragmentation of the valence quarks in the projection is given by
the simple formula for the rapidity distribution of interactions
with gluon in the target~\cite{Mehtar-Tani1,Mehtar-Tani2}

\vspace{-0.3cm}

\begin{equation}  %%% Eq.3
       \frac{dN}{dy}=\frac{C}{(2\pi
       )^2}\int{\frac{d^2p_{T}}{{p_{T}}^2}}x_{1}q_{v}(x_{1})\varphi (x_{2},p_{T})
\label{eq:eq3} %Eq.3
\end{equation}

\noindent here $x_{1}q_{v}(x_{1})$ is the valence quark distribution
of a nucleus, $\varphi (x_{2},p_{T})$ is the gluon distribution
of another nucleus, $p_{T}$ is the transverse momentum of the
produced quark and $y$ its rapidity. One important prediction of the
gluon saturation with geometric scaling is the geometric scaling:
the gluon distribution depends on $x$ and $p_{T}$ only through the
scaling variable $p_{T}^2/Q_{s}^2(x)$, here
$Q_{s}^2(x)=A^{1/3}Q_{0}^2x^{-\lambda}$.Geometric scaling was
confirmed experimentally at HERA~\cite{Breitweg}.The fit value
$\lambda=0.2-0.3$ agrees with theoretical estimates based on
next-to-leading order Balitskii-Fadin-Kuraev-Lipatov (BFKL)
results~\cite{Kuraev,Ya,Lipatov}.

The longitudinal momentum fractions carried, respectively, by the
valence quark in the projectile and the soft gluon in the target are
$x\equiv x_{1}=(p_{T}/\sqrt{s})exp(y)$ , and
$x_{2}=(p_{T}/\sqrt{s})exp(-y)$. Then we can deduce the relation as
follows~\cite{Mehtar-Tani1,Mehtar-Tani2}

\vspace{-0.3cm}

\begin{equation}  %%% Eq.4
      x\equiv x_{1},\ x_{2}=xe^{-2y},\ p_{T}^2=x^2se^{-2y}
\label{eq:eq4} %Eq.4
\end{equation}

\noindent The gluon distribution is defined as

\vspace{-0.3cm}

\begin{equation}  %%% Eq.5
      \varphi (x_{2},p_{T})= \varphi
      (\frac{p_{T}^2}{Q_{s}^2(x_2)})=4\pi\cdot \frac{p_{T}^2}{Q_{s}^2(x_2)}\cdot exp({-\frac{p_{T}^2}{Q_{s}^2(x_2)}})
\label{eq:eq5} %Eq.5
\end{equation}

When we use the variables as Eq.4, we can take the relation as
follows

\vspace{-0.3cm}

\begin{equation}  %%% Eq.6
       \frac{p_{T}^2}{Q_{s}^2(x_2)}=\frac{(x\cdot \sqrt{s}\cdot
       e^{-y})^2}{A^{1/3}Q_{0}^2x_2^{-\lambda}}=\frac{x^{2+
       \lambda}}{e^{2(1+\lambda)y}}\cdot \frac{s}{Q_0^2}\cdot
       \frac{1}{A^{1/3}}
\label{eq:eq6} %Eq.6
\end{equation}

\noindent If we make $\frac{p_{T}^2}{Q_{s}^2(x_2)}=x^{2+
       \lambda}\cdot e^{\tau}$£¬from Eq.6 we can take

\vspace{-0.3cm}

\begin{equation}  %%% Eq.7
       e^{\tau}=\frac{1}{e^{2(1+\lambda)y}}\cdot \frac{s}{Q_0^2}\cdot
       \frac{1}{A^{1/3}}
\label{eq:eq7} %Eq.7
\end{equation}

\noindent and a geometrical scaling with rapidity is introduced as

\vspace{-0.3cm}

\begin{equation}  %%% Eq.8
       \tau=ln(\frac{s}{Q_0^2})-lnA^{1/3}-2(1+\lambda)y
\label{eq:eq8} %Eq.8
\end{equation}

\noindent Thus Eq.3 is given as follows

\begin{equation}  %%% Eq.9
       \frac{dN}{dy}=\frac{C}{2\pi
       }\int_0^1\frac{dx}{x}xq_{v}(x)\varphi (x^{2+\lambda},e^{\tau})
\label{eq:eq9} %Eq.9
\end{equation}

\noindent As mentioned before, the dependence of $\tau$ on $y$ in
Ref.[14,15] is related to the peak position, but here we use the
dependence of $\tau$ on $y$ to define the rapidity limit of gluon
saturation. A rapidity variable is introduced in Eq.8 to discuss the
gluon saturation rapidity region. The rapidity limit variable $y$ is

\vspace{-0.5cm}

\begin{equation}  %%% Eq.10
       y=\frac{1}{1+\lambda}(y_b-lnA^{1/6})+\frac{1}{2(1+\lambda)}(ln \frac{m_n^2}{Q_0^2}-\tau)
\label{eq:eq10} %Eq.10
\end{equation}

\noindent here $y_b\thickapprox ln(\sqrt{s}/m_n)$ is the beam
rapidity with nucleon mass $m_n$. It usually defines $x\leqslant
0.01$ as a small $x$ region, and in this region, it is taken as the
gluon saturation region. By taking $\tau=ln(1/x)|_{x=0.01}$ , we can
figure out the rapidity region of gluon saturation as follows

\vspace{-0.5cm}

\begin{equation}  %%% Eq.11
        \begin{split}
           &y_{saturation}\\
           &=\frac{1}{1+\lambda}(y_b-lnA^{1/6})+\frac{1}{2(1+\lambda)}(ln \frac{m_n^2}{Q_0^2}-\tau|_{x=0.01})
        \end{split}
\label{eq:eq11} %Eq.11
\end{equation}

\noindent The gluon distribution is

\vspace{-0.5cm}

\begin{equation} %%% Eq.12
        \varphi (x_{2},p_{T})=\left\{ \begin{array}{ll}
                              \hspace{-0.1cm}const & \textrm{$(x<0.01)$}\\
                             \hspace{-0.1cm}4\pi\frac{p_{T}^2}{Q_{s}^2(x_2)}\cdot
                              e^{-\frac{p_{T}^2}{Q_{s}^2(x_2)}}\cdot
                              (1-x_2)^4 & \textrm{$(x>0.01)$}
                              \end{array}\right.
\label{eq:eq12}  %Eq.12
\end{equation}

\noindent The valence quarks distribution is

\vspace{-0.5cm}

\begin{equation}  %%% Eq.13
       xq_v(x)=\left\{ \begin{array}{ll}
                              \propto \ x^{0.5} & \textrm{$(x<0.01)$}\\
                              \propto \ (xu_v+xd_v) & \textrm{$(x>0.01)$}
                              \end{array}\right.
\label{eq:eq13} %Eq.13
\end{equation}

\noindent So the net-baryon distribution which originates from the
projectile is

\vspace{-0.5cm}

\begin{equation}  %%% Eq.14
       \frac{dN}{dy}=\left\{ \begin{array}{ll}
                       \hspace{-0.1cm}\propto \ exp\{\frac{1+\lambda}{2+\lambda}y\} \qquad \qquad \quad(0\leq y\leq y_{saturation}) & \textrm{$$}\\
                       \\[0.1cm]
                       \hspace{-0.1cm}\propto \int_{0.01}^1\frac{dx}{x}(xu_v+xd_v)4\pi \cdot \frac{p_{T}^2}{Q_{s}^2(x_2)}\\
                       \quad \cdot exp\{-\frac{p_{T}^2}{Q_{s}^2(x_2)}\}\cdot
                       (1-x_2)^4 \quad (y>y_{saturation}) & \textrm{$$}
                             \end{array}\right.
\label{eq:eq14} %Eq.14
\end{equation}

The contribution of valence quarks in the other beam nucleus is
added incoherently by changing $y\rightarrow -y$.  The total
rapidity distributions of the symmetry interaction systems are the
summation of the contributions from the projectile and target,
respectively

\begin{equation}  %%% Eq.15
        \frac{dN}{dy}|_{total}= \frac{dN}{dy}(y)+ \frac{dN}{dy}(-y)
\label{eq:eq15} %Eq.15
\end{equation}

Compared with Ref.[14,15], we will introduce effective dynamic
quark mass in this paper by substituting the transverse momentum
$p_{T}$ as $\sqrt{(x\sqrt{s}e^{-y})^2-m^2}$ , i.e.
$p_{T}=\sqrt{(x\sqrt{s}e^{-y})^2-m^2}$ £¬here $m$ is the dynamic
quark mass, which was given by baryon mass in Ref.[14,15]. The
detailed study of dynamic mass with collision energies will be given
in the next section.

\section{The net baryon distributions in the SPS to LHC energy regions}
The integrated net-proton rapidity distributions are scaled by a
factor of 2.05~\cite{Iancu} to obtain the net-baryon distributions.
The rapidity distributions of net baryon for different energies of
relativistic heavy nuclear collisions (Pb + Pb and Au + Au) at SPS
and RHIC are given in Figure 1.

\begin{figure}[h!]
\centering \resizebox{0.45\textwidth}{!}{
\includegraphics{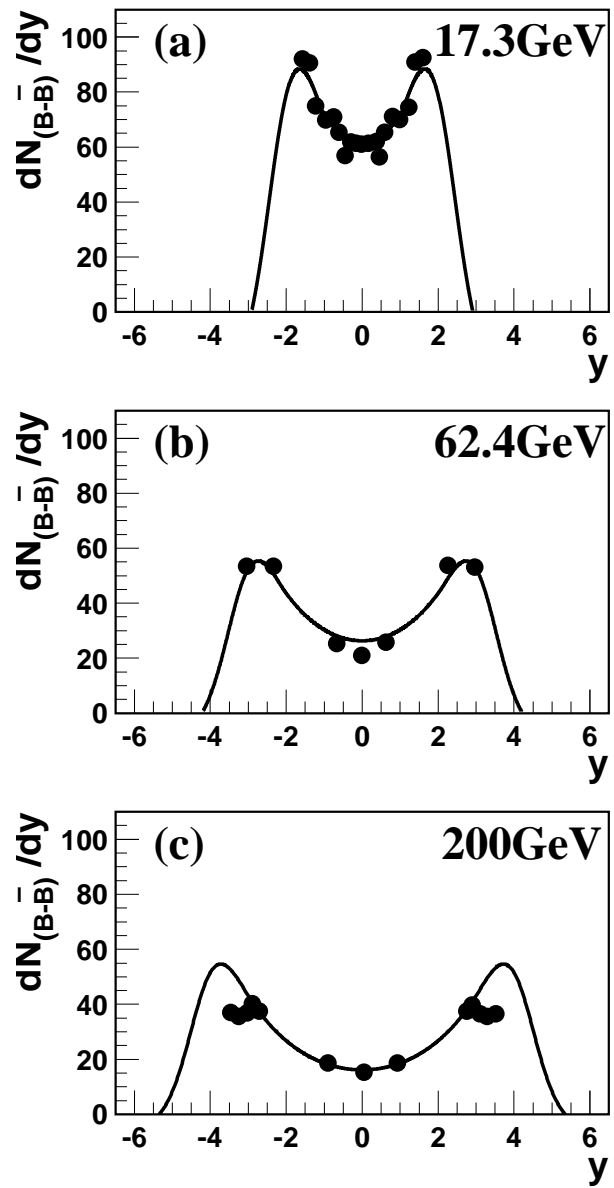}}
\caption{Net-Baryon distributions of central
collisions at SPS $\sqrt{s_{NN}}=17.3$ GeV of Pb-Pb interactions
and at RHIC $\sqrt{s_{NN}}=62.4$ and $200$ GeV of Au-Au interactions; the experimental results are from [3-9]. }
\label{fig1}
\end{figure}

The estimated numbers of participants are $390$, $315$, and $357$
for $\sqrt{s_{NN}}=17.3$, $62.4$, and $200$ GeV, respectively. The
solid circles correspond to the experimental result of central
collisions~\cite{Arsene,Bearden1,Klay,Bearden2,Barrette,Ahle,Appelshauser},
and the real lines are the calculated results from the gluon
saturation model with geometric scaling. It is found that our model
describes the experimental data of net-baryon distributions very
well when we discuss Pb-Pb center collisions at the SPS energy
region and Au-Au center collisions at the RHIC energy region. The
$\lambda =0.2$ and $Q_{0}^{2}=0.05$ GeV$^2$ are fixed in our whole
calculations.

\begin{figure}[h!]
\centering \resizebox{0.45\textwidth}{!}{
\includegraphics{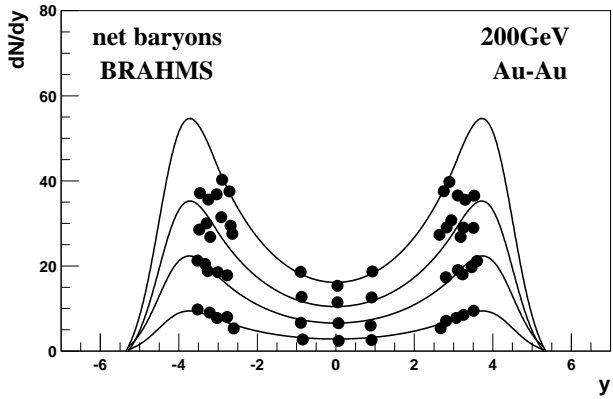}}
\caption{Rapidity distributions of net baryons at
different centrality; the experimental results are from [27], and
the real lines are from our model. }
\label{fig2}
\end{figure}

We show in Fig.2 the computation resulting from our discussions, the
centrality dependence of the rapidity distribution at
$\sqrt{s_{NN}}=200$ GeV. The estimated numbers of participants are
280, 200, 114 and 54 for centralities of $0\%-10\%$, $10\%-20\%$,
$20\%-40\%$, and $40\%-60\%$(from top to bottom) in Au+Au collisions at
RHIC energies of $\sqrt{s_{NN}}=200$ GeV, respectively.

By studying the experimental results with our model, we may get the
conclusion for the net-baryon distribution from SPS to LHC as
follows:
£¨1£©The gluon saturation model with geometric scaling may
be a good theory to study net-baryon distribution at central
collisions at SPS and RHIC. The contributions of spectator nucleons
can be neglected when considering only the central collisions. The
gluon saturation feature of central rapidity can be studied from our
discussion.  The values of central rapidity of gluon saturation
($y_{saturation}$) are 1.06, 2.013 and 3.10 from Eq.11 for
$\sqrt{s_{NN}}= 17.3$, $62.4$, and $200$ GeV, respectively.

\begin{figure}[h!]
\centering \resizebox{0.45\textwidth}{!}{
\includegraphics{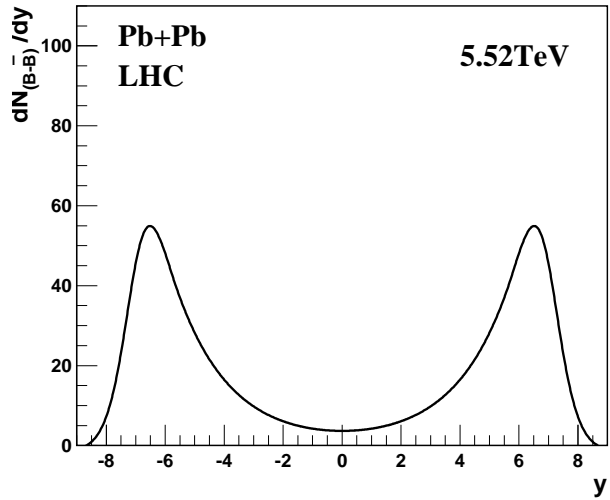}}
\caption{The rapidity distribution of net baryons in
central Pb+Pb collisions at LHC energies of $\sqrt{s_{NN}}=
5.52$TeV. The theoretical distribution is from our discussion. }
\label{fig3}
\end{figure}

(2)  The net baryon rapidity distribution in central Pb+Pb
collisions at LHC energies of  $\sqrt{s_{NN}}= 5.52$ TeV is predicted
by gluon saturation model with geometric scaling. The theoretical
distribution is shown in Fig.3 for $y_{saturation}=5.86$. The gluon
saturation region is larger than those of RHIC and SPS. It is found
that the separation of two symmetric peaks of net-baryon is much
wider than that of SPS and RHIC.

\begin{figure}[h!]
\centering \resizebox{0.45\textwidth}{!}{
\includegraphics{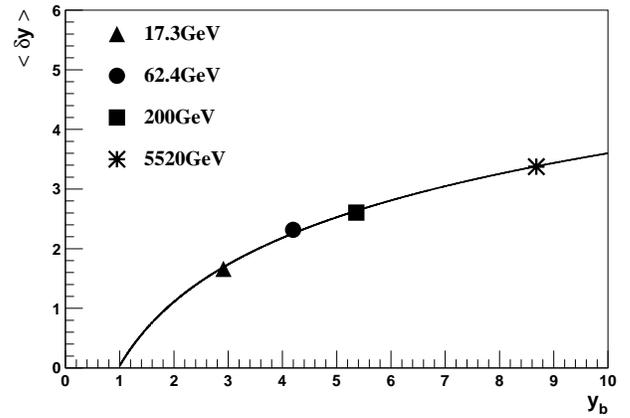}}
\caption{The dependence of mean rapidity loss
$<\delta y>$ on beam rapidity $y_b$. The $\blacktriangle$ ,
$\bullet$, $\blacksquare$, $\ast$ are the calculated results by our
discussion for $\sqrt{s_{NN}}= 17.3$, 62.4, 200 and $5520$ GeV,
respectively. The real curve is the fit curve with $<\delta
y>=1.548\cdot\ln(y_b)+0.036$ }
\label{fig4}
\end{figure}

(3) The mean rapidity loss $<\delta y>=y_{b}-<y>$ is shown in Fig.
4. We show that the dependence of mean rapidity loss increase on
$y_b$ as

\vspace{-0.5cm}

\begin{equation}  %%% Eq.16
       <\delta y>=1.548\cdot\ln(y_b)+0.036
\label{eq:eq16} %Eq.16
\end{equation}

In Figure.4, the star ($\ast$) is our prediction result of the mean
rapidity loss $<\delta y>$ for Pb + Pb central collisions at LHC
energies of $\sqrt{s_{NN}}= 5.52$ TeV ($y_b=8.68$). In high-energy
central nucleus-nucleus collisions, the baryon matter will be slowed
down and will lose a few units of rapidity. The term "nuclear
stopping power" was introduced in high-energy nucleus-nucleus
collisions by Busza and Goldhaber to refer to the degree of stopping
an incident nucleon suffers when it impinges on the nuclear matter
of another nucleus. Usually the mean rapidity loss $ <\delta
y>=y_{b}-<y>$ was used to represent the nuclear stopping power of
nucleus -nucleus collisions.

\begin{figure}[h!]
\centering \resizebox{0.45\textwidth}{!}{
\includegraphics{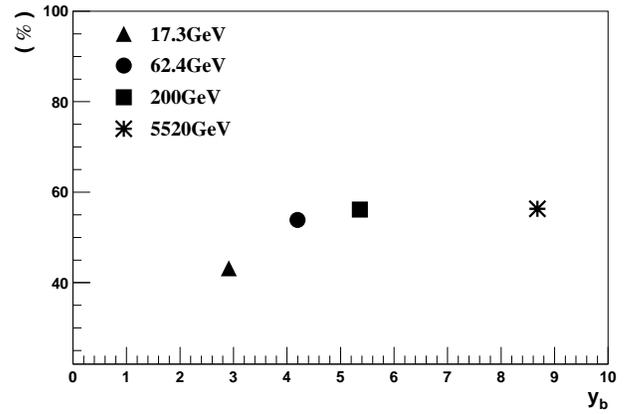}}
\caption{The dependence of the percentage of
net-baryon from the central gluon saturation region on the colliding
energies at $\sqrt{s_{NN}}=17.3$, 62.4 , 200 and 5520 GeV. }
\label{fig5}
\end{figure}

(4) It is shown in Fig.5 that the dependence of percentage ratios of
the net-baryon production from the central gluon saturation region
on incident energy at $\sqrt{s_{NN}}=17.3$, 62.4 , 200 and 5520 GeV.
It is found that the percentage ratio from the central gluon
saturation region increases
 with the increase of the colliding energies. The percentage ratio from
 the saturation region rises rapidly from SPS $\sqrt{s_{NN}}=17.3$GeV to RHIC $\sqrt{s_{NN}}=62.4$GeV,
 but slowly from RHIC $\sqrt{s_{NN}}=200$ GeV to LHC $\sqrt{s_{NN}}=5520$ GeV. It is found that the percentage
 ratio is $43.17\%$ at $\sqrt{s_{NN}}=17.3$GeV, but $56.31\%$ at $\sqrt{s_{NN}}=5520$ GeV. It seems that
 that the number of more than half of the net baryon at LHC comes from
 the central gluon saturation region.

\begin{figure}[h!]
\centering \resizebox{0.45\textwidth}{!}{
\includegraphics{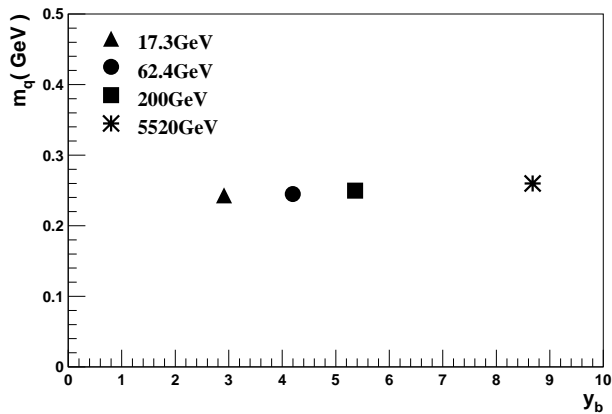}}
\caption{The dependence of the effective quark mass
on incident projective rapidity in the CMS  at $\sqrt{s_{NN}}=17.3$,
62.4 , 200 and 5520 GeV. }
\label{fig6}
\end{figure}

£¨5£©The dependence of the effective quark mass $m_q$ on incident
projective rapidity in the CMS  at $\sqrt{s_{NN}}=17.3$, 62.4 , 200
and 5520 GeV is shown in Fig.6. It is found that the effective quark
mass $m_q$ varies slowly with the varying of the incident energies,
and $m_q$ varies among 0.24 to 0.26.

\section{Summary and conclusion}

As discussed in Ref. [14,15],  two distinct and symmetric peaks with
respect to rapidity $y$ occur at SPS energies and beyond in A + A
collisions. A geometrical scaling feature of peak position of net
baryon rapidity distributions was proposed to discuss about the
net-baryon distribution~\cite{Mehtar-Tani1,Mehtar-Tani2}. The
rapidity separation between the peaks increases with energy and
decreases with the increasing mass number, $A$, reflecting larger
baryon stopping for heavier nuclei, as was investigated
phenomenologically in the Non-uniform Flow
Model(NUFM)~\cite{Feng1,Feng2,Feng3,Feng4}. In this work we show the
geometrical scaling with gluon saturation rapidity limit, and also
discuss the net-baryon rapidity distribution feature in the
SPS£¬RHIC and LHC.

 A saturation model for net-baryon distributions that successfully
 describes the net-baryon rapidity distributions and their energy
 dependence is presented in this paper.  The remarkable feature of
 geometric scaling predicted by our discussion is reflected in the
 net-baryon rapidity distribution, providing a direct test of gluon
 saturation rapidity and $x$ regions. The gluon saturation model is
 proposed by introducing a rapidity variable with gluon saturation
 region to define the gluon saturation region of central rapidity
 region of centrally colliding heavy ions at ultra-relativistic energies.

The gluon saturation features of central rapidity at SPS and RHIC
can be investigated. It is found that the values of central rapidity
of gluon saturation region increase with colliding energy. The
detailed dependence of rapidity ($y_{saturation}$) of central gluon
saturation on colliding energy is also investigated in this paper.
We also predict the net baryon rapidity distribution in central
Pb+Pb collisions at LHC energies of $\sqrt{s_{NN}}=5.52TeV$ by gluon
saturation model with geometric scaling. The gluon saturation region
is larger than those of RHIC and SPS, and the separation of two
symmetric peaks of net-baryon is much wider than that of SPS and
RHIC.

It is shown that gluon saturation feature is an important feature
with the increasing of colliding energy. It seems that more than
half of the produced net baryon numbers ($56.31\%$) at LHC come from
the central gluon saturation region, but the percentage ratio is
$43.17\%$ at SPS $\sqrt{s_{NN}}=17.3$GeV.

The dependencies of the percentage of net-baryon from the central
gluon saturation region and stopping power on colliding energies are
also studied in this paper. It is also found that the mean rapidity
loss shows a linear dependence on $\ln(y_b)$. From that we can
predict the mean rapidity loss for future LHC experimental data.

\section{Acknowledgments}
This work was supported by National Natural Science Foundation of China (10975091),
Excellent Youth Foundation of Hubei Scientific Committee (2006ABB036)and Education Commission of Hubei
Province of China (Z20081302).The authors is indebted to
Prof. Lianshou Liu for his valuable discussions and very helpful
suggestions.

{}

\end{document}